\title{Integrated Interleaved Codes as Locally Recoverable Codes:
Properties and Performance}
\author{Mario Blaum and Steven R. Hetzler\\
IBM Almaden Research Center\\
San Jose, CA 95120
}
\documentclass[12pt]{article}
\usepackage{latexsym}
\usepackage{multirow}
\usepackage{amsmath}
 \newtheorem{theo}{Theorem}[section]

 \newtheorem{defin}{Definition}[section]
 \newtheorem{ex}{Example}[section]
 
 \newtheorem{cor}{Corollary}[section]
 
\newtheorem{COROLLARY}{\indent Corollary}

\newtheorem{EXAMPLE}{\indent Example}

\newtheorem{THEOREM}{\indent Theorem}

\newtheorem{REMARK}{\indent Remark}

\newcommand{\ga}{\mbox{$\gamma$}}

\newcommand{\fullstop}{\hspace{-0.85em} {\bf .}}

\newcommand{\uv}{\mbox{$\underline{v}$}}

\newcommand{\us}{\mbox{$\underline{s}$}}
\newcommand{\uu}{\mbox{$\underline{u}$}}

\newcommand{\uc}{\mbox{$\underline{c}$}}

\newcommand{\hs}{\mbox{$\hat{s}$}}

\newcommand{\ra}{\mbox{$\rightarrow$}}

\newcommand{\al}{\mbox{$\alpha$}}
\newcommand{\si}{\mbox{$\sigma$}}
\newcommand{\eq}{\mbox{$\, =\,$}}

\newcommand{\tr}{\mbox{$\triangle$}}
\newcommand{\qed}{\hfill$\Box$\\[1ex]}
\newcommand{\pf}{{\bf Proof: }}

\newcommand{\uzero}{\mbox{$\underline{0}$}}

\newcommand{\xor}{\mbox{$\,\oplus\,$}}

\newcommand{\C}{\mbox{${\cal C}$}}

\newcommand{\cO}{\mbox{${\cal O}$}}

\newcommand{\br}{\\ }
\newcommand{\ce}{\begin{center}}
\newcommand{\cen}{\end{center}}

\newcommand{\ipb}{\begin{description}}
\newcommand{\ipn}{\end{description}}

\newcommand{\qb}{\begin{quote}}
\newcommand{\qn}{\end{quote}}

\newcommand{\tp}{\begin{titlepage}}
\newcommand{\tpn}{\end{titlepage}}
\newcommand{\zb}{\begin{figure}[hbtp]}
\newcommand{\zn}{\end{figure}}

\newcommand{\EQX}[1]{\begin{equation}\label{#1}}
\newcommand{\ENX}{\end{equation}}
\newcommand{\EQL}{\begin{eqnarray*}}
\newcommand{\ENL}{\end{eqnarray*}}
\newcommand{\EQLX}[1]{\begin{eqnarray}\label{#1}}
\newcommand{\ENLX}{\end{eqnarray}}

\newcommand{\open}{\begin{document}}
\newcommand{\close}{\end{document}}
\newcommand{\lfcr}[1]{\br\hspace*{#1em}}
\newenvironment{mat}[1]
{\left[\begin{array}{#1}}{\end{array}\right]}
\newcommand{\GAMMA}{\Gamma}
\newcommand{\DELTA}{\Delta}
\newcommand{\THETA}{\Theta}
\newcommand{\LAMBDA}{\Lambda}
\newcommand{\XI}{\Xi}
\newcommand{\PI}{\Pi}
\newcommand{\SIGMA}{\Sigma}
\newcommand{\UPSILON}{\Upsilon}
\newcommand{\PHI}{\Phi}
\newcommand{\PSI}{\Psi}
\newcommand{\OMEGA}{\Omega}
\newcommand{\bldgreek}[1]{\mbox{\boldmath $#1$}}
\newcommand{\bldbeta}{\bldgreek{\beta}}
\newcommand{\bldgamma}{\bldgreek{\gamma}}
\newcommand{\blddelta}{\bldgreek{\delta}}
\newcommand{\bldepsilon}{\bldgreek{\epsilon}}
\newcommand{\bldvarepsilon}{\bldgreek{\varepsilon}}
\newcommand{\bldzeta}{\bldgreek{\zeta}}
\newcommand{\bldeta}{\bldgreek{\eta}}
\newcommand{\bldtheta}{\bldgreek{\theta}}
\newcommand{\bldvartheta}{\bldgreek{\vartheta}}
\newcommand{\bldiota}{\bldgreek{\iota}}
\newcommand{\bldkappa}{\bldgreek{\kappa}}
\newcommand{\bldlambda}{\bldgreek{\lambda}}
\newcommand{\bldmu}{\bldgreek{\mu}}
\newcommand{\bldnu}{\bldgreek{\nu}}
\newcommand{\bldxi}{\bldgreek{\xi}}
\newcommand{\bldpi}{\bldgreek{\pi}}
\newcommand{\bldvarpi}{\bldgreek{\varpi}}
\newcommand{\bldrho}{\bldgreek{\rho}}
\newcommand{\bldvarrho}{\bldgreek{\varrho}}
\newcommand{\bldsigma}{\bldgreek{\sigma}}
\newcommand{\bldvarsigma}{\bldgreek{\varsigma}}
\newcommand{\bldtau}{\bldgreek{\tau}}
\newcommand{\bldupsilon}{\bldgreek{\upsilon}}
\newcommand{\bldphi}{\bldgreek{\phi}}
\newcommand{\bldvarphi}{\bldgreek{\varphi}}
\newcommand{\bldchi}{\bldgreek{\chi}}
\newcommand{\bldpsi}{\bldgreek{\psi}}
\newcommand{\bldomega}{\bldgreek{\omega}}
\begin{document}
\parindent=10pt
\maketitle
\begin{abstract}
Considerable interest has been paid in recent literature to codes combining
local and global properties for erasure correction. Applications are
in cloud type of implementations, in which fast recovery of a failed
storage device is important, but additional protection is required in
order to avoid data loss, and in RAID type of architectures, in which
total device failures coexist with silent failures at the page or
sector level in each device. Existing solutions to these problems
require in general relatively large finite fields. The techniques of
Integrated Interleaved Codes (which are closely related to
Generalized Concatenated Codes) are proposed to reduce significantly
the size of the finite field, and it is shown that when the
parameters of these codes are judiciously chosen, their performance
may be competitive with the one of codes optimizing the minimum distance.

\vspace{.3cm}

\end{abstract}

\section{Introduction}
\label{Introduction}

In recent literature there was considerable interest in obtaining
codes with local and global properties for erasure correction. The
idea is to divide data symbols into sets and add parity symbols
(local parities) to
each set independently (preferably, using and MDS code). So, in case
a number of erasures not exceeding the
number of parity symbols occurs in a set, such erasures are rapidly
recovered. In addition to the local parities, a number of
global parities are added. Those global parities involve all of the
data symbols, and may include the local parity symbols. The
idea is that the global parities can correct situations in which the
erasure-correcting power of the
local parities has been exceeded. 

\begin{figure}
$$
\begin{array}{|c|c|c|c||c|}
\hline
{\bf D}&{\bf D}&{\bf D}&{\bf D}&{\bf L}\\
\hline
{\bf D}&{\bf D}&{\bf D}&{\bf D}&{\bf L}\\
\hline
{\bf D}&{\bf D}&{\bf D}&{\bf D}&{\bf L}\\
\hline
{\bf D}&{\bf D}&{\bf G}&{\bf G}&{\bf L}\\
\hline
\end{array}
$$
\caption{\label{fig1}
An example of placement of data and local and global parities}
\end{figure}

The situation is illustrated in
Figure~\ref{fig1} showing a $4\times 5$ array, in which each row of
data (denoted by ${\bf D}$) is encoded with a local parity (denoted by
${\bf L}$). That is, each local parity ${\bf L}$ affects only the row where it
belongs. The two global parities, denoted by ${\bf G}$, affect all the
data, and possibly (but not necessarily) the local parities ${\bf L}$.
Similarly, the local parities ${\bf L}$ affect the data in the
corresponding row, and they may or may not extend to the global
parities ${\bf G}$. It is possible for the rows to have different
lengths, one may simply assume that some of the data symbols ${\bf D}$ are
zero (the process known as shortening of a code~\cite{ms}).

An example of a code with local and global properties in which those
global and local parities are independent from each other is provided
in~\cite{sa}, as illustrated in Figure~\ref{fig2}.

\begin{figure}
$$
\begin{array}{ccc}
\begin{array}{|c|}
\hline
{\bf L}\\
\hline
{\bf L}\\
\hline
\end{array}
&
\begin{array}{|c|c|c|c|c|}
\hline
{\bf D}&{\bf D}&{\bf D}&{\bf D}&{\bf D}\\
\hline
{\bf D}&{\bf D}&{\bf D}&{\bf D}&{\bf D}\\
\hline
\end{array}
&
\begin{array}{|c|c|c|c|}
\hline
{\bf G}&{\bf G}&{\bf G}&{\bf G}\\
\hline
\end{array}
\end{array}
$$
\caption{\label{fig2}
The Xorbas code with ten data, two local and four
global symbols}
\end{figure}

In Figure~\ref{fig2}, the 10 data symbols are divided into two sets,
and each of these sets is protected with a (local) parity symbol. In
addition, the 10 data symbols are protected using 4 extra (global)
parity symbols corresponding to a Reed-Solomon (RS) code~\cite{ms} over
$GF(16)$. This code has locality 5 on data, which means, if
just one data symbol is erased, such symbol can be recovered by using
the remaining 5 symbols in the parity set. There is no locality for
the global symbols, but the approach is letting those symbols remain
erased, at the price of reducing the erasure-correcting capability of
the RS code. When more than one erasure occurs in a parity set, the
RS code is invoked. It is easy to see that this code has minimum
distance 5, i.e., any four erasures can be recovered.

The interest in erasure correcting codes with local and global
properties arises mainly from
two applications. One of them is the cloud. A cloud configuration
may consist of many storage devices, of which some of them may even
be in different geographical locations and the data is distributed
across them. In the case that one or more of those devices fails, it is
desirable to recover its contents ``locally,'' that is, using a few
parity devices within a set of limited size in order to affect
performance as little as possible. However, the local parity may not
be enough. We want extra
protection in case the erasure-correcting capability of a local set is
exceeded: in that case, some devices containing global parities are
incorporated, and when the local correction power is exceeded,
the global parities
are invoked and correction is attempted. If such a situation occurs,
there will be an impact on performance,
but data loss may be averted. It is expected that the cases in which
the local parity is
exceeded are relatively rare events, so the aforementioned impact on
performance does not occur frequently.
As an example of this type of application, we refer the reader to the
description of the Azure system~\cite{hsx} or to the Xorbas code discussed
in Figure~\ref{fig2}~\cite{sa}.

A second application occurs in the context of Redundant Arrays of
Independent Disk (RAID) architectures~\cite{g}. In this case, a RAID
architecture protects against one or more storage device failures.
For example, RAID 5 adds one extra parity device, allowing for the
recovery of the contents of one failed device, while RAID 6 protects
against up to two device failures. In particular, if those devices
are Solid State Drives (SSDs), like flash memories, their
reliability decays with time and with the number of writes and
reads~\cite{M}.
The information in SSDs is generally divided into pages, each page
containing its own internal Error-Correction Code (ECC). It may
happen that a particular page degrades and its ECC is exceeded.
However, this situation may not be known to the user until the page is
accessed (what is known as a silent failure). Assuming an SSD has
failed in a RAID~5 scheme, if during reconstruction a silent page
failure is encountered in one of the surviving SSDs, then data loss
will occur. A method around this situation is using RAID~6. However,
this method is costly, since it requires two whole SSDs as parity. It is more
desirable to divide the information in a RAID type of architecture
into $m\times n$ stripes: $m$ represents the size of a stripe, and
$n$ is the number of SSDs. The RAID architecture can be viewed as
consisting of a large number of stripes, each stripe encoded and
decoded independently. Certainly, codes like the ones used in cloud
applications can be used as well for RAID applications, like the one
depicted in
Figure~\ref{fig1} describing a $4\times 5$ stripe with two global
parities. It has better rate than RAID~6, which would require two
whole columns devoted to parity. Of course, the choice of code
depends on the
statistics of errors and on the frequency of silent page failures.

From now on, we call symbols the entries of a code with local and
global properties. Such symbols can be whole devices (for example, in
the case of cloud applications) or pages (in the case of RAID
applications for SSDs).

Let us follow the notation in~\cite{b}. We will consider codes as consisting
of $m\times n$
arrays, such that each row in an array contains $\ell$ local parities. In
the usual coding notation~\cite{ms}, each row corresponds to an
$[n,n-\ell]$ code. We will further assume that
the local codes are MDS,
i.e., up to $\ell$ erasures in each row may be corrected locally by
invoking the non-erased symbols in the row. In
addition, a number $g$ of global parities are added to the code.
Let us state the definition of Locally Recoverable Codes formally.

\begin{defin}
\label{defLRC}
{\em 
Consider a code $\C$ over a finite field $GF(q)$ consisting of
$m\times n$ arrays such that, given integers $\ell$ and $g$
where $1\leq\ell\,<\,n$ and $0\leq g\,<\,m(n-\ell)$, the arrays
satisfy: 

\begin{enumerate}

\item Each row in each array in $\C$ is in an
$[n,n-\ell,\ell +1]$ MDS code over $GF(q)$.

\item Reading the symbols of $\C$ row-wise, $\C$ is an
$[mn,m(n-\ell)-g]$ code over $GF(q)$. 

\end{enumerate}

Then we say that $\C$ is an $(m,n;\ell,g)$ Locally Recoverable (LRC) Code.\qed
}
\end{defin}

In Definition~\ref{defLRC}, each row corresponds to a parity set.
Strictly speaking, it is not necessary that the
parity sets as given by rows in this description are disjoint. For
example, Definition~1 of local-error correction (LEC) codes
in~\cite{pklk} does not
make this assumption; however, most constructions do (see for
example~\cite{tb}, to be discussed below). 

Certainly, the number of global parities in Definition~\ref{defLRC}
may well be $g\eq 0$. So, an $(m,n;\ell,0)$ LRC code would correspond
to a RAID scheme~\cite{g} 
in which each row is protected against up to $\ell$ erasures (in
particular, $(m,n;1,0)$ corresponds to RAID~5 and $(m,n;2,0)$
corresponds to RAID~6).

So, the question is, how do we add the $g$ global parities to the
array such that the code is optimized? There are several possible criteria
for optimization in literature, so let us briefly review them.

Given an $(m,n;\ell,g)$ LRC code $\C$, a natural
place to start is with the minimum distance $d$ of the code. A
Singleton type of bound on $d$ was obtained in~\cite{ghsy}\cite{pklk}, which we
present next adapted to Definition~\ref{defLRC}.
Denoting by $\lfloor x\rfloor$ the floor of $x$, the minimum distance
$d$ of $\C$ is bounded by

\begin{eqnarray}
\label{dist}
d&\leq&\ell +g+\ell\left\lfloor{g\over n-\ell}\right\rfloor +1.
\end{eqnarray}

An important subcase of bound~(\ref{dist}) occurs when
$\ell+g\,<\,n$. In this case, bound~(\ref{dist}) becomes simply

\begin{eqnarray}
\label{dist1}
d&\leq & \ell + g+1.
\end{eqnarray}

For example, Figure~\ref{fig1} depicts a (4,5;1,2) LRC code.
Bound~(\ref{dist1}) states that for this code $d\leq 4$. This is
easy to see, since four erasures in the same row are uncorrectable:
there are not enough parities, we have only three. The argument to prove
the more general bound~(\ref{dist}) proceeds similarly.

Although we will next see stronger criteria
for optimality of LRC codes, we follow the traditional denomination
in literature and we call the LRC codes meeting bound~(\ref{dist})
optimal LRC codes.
Most of the work on LRC codes concentrates on constructing optimal LRC codes
(see~\cite{hcl}\cite{kna}\cite{pd}\cite{pklk}\cite{rk}\cite{sa}\cite{sd}\cite{tb}\cite{wz}
and references 
within). Bound~(\ref{dist}), being a Singleton type of
bound, does not take into account the size of the field (for a bound
that does consider the size of the field, see~\cite{cm}).
Certainly, it is desirable to have a field as small as possible. In
some of the early constructions, the field is relatively large, but a
satisfactory solution to the problem of constructing optimal LRC
codes is given
in~\cite{tb}, where the size of the field is at least $mn$, i.e., the
length of the code, as is the case with RS codes (in fact, the
construction in~\cite{tb} can be viewed as a generalization of RS
codes, which correspond to the
special case $m\eq 1$). 

A second (and stronger) approach to optimizing LRC codes is given by
Partial MDS (PMDS)
codes \cite{bhh}\cite{bpsy}\cite{ghjy}\cite{hsx}
(in~\cite{ghjy}\cite{hsx}, PMDS codes are called Maximally
Recoverable codes).
An $(m,n;\ell,g)$ PMDS code, in addition to correcting up to $\ell$
erasures per row, allows for the correction of
$g$ erasures anywhere. Another way of stating the PMDS property, is
that a punctured code~\cite{ms}
consisting of puncturing any $\ell$ locations in each row, is
an $[m(n-\ell),m(n-\ell)-g,g+1]$ code, i.e., it is an MDS code.

\begin{figure}
$$
\begin{array}{cc}
\begin{array}{|c|c|c|c|c|}
\hline
X&\phantom{X}&\phantom{X}&\phantom{X}&\phantom{X}\\
\hline
\phantom{X}&X&\phantom{X}&\phantom{X}&X\\
\hline
\phantom{X}&\phantom{X}&\phantom{X}&\phantom{X}&X\\
\hline
\phantom{X}&\phantom{X}&X&X& \\
\hline
\end{array}
&
\begin{array}{|c|c|c|c|c|}
\hline
X&\phantom{X}&\phantom{X}&\phantom{X}&\phantom{X}\\
\hline
\phantom{X}&X&\phantom{X}&X&X\\
\hline
\phantom{X}&\phantom{X}&\phantom{X}&\phantom{X}&X\\
\hline
\phantom{X}&\phantom{X}&\phantom{X}&\phantom{X}&X\\
\hline
\end{array}
\end{array}
$$
\caption{\label{fig3}
Patterns that can be corrected by a (4,5;1,2) PMDS code }
\end{figure}

Figure~\ref{fig3} illustrates the correction power of a
(4,5;1,2) PMDS code.
The array on the left, once the first and the
third rows are corrected using the local parity, can correct two
erasures in the second row and two erasures in the fourth row. The
array on the right, once the first, the third and the fourth rows are
corrected, can correct three erasures in the second row. This
(4,5;1,2) PMDS code
has minimum distance 4, since 4 erasures in the same row cannot be
corrected. An optimal LRC with the same parameters has also minimum
distance 4. However, an optimal LRC code in general cannot correct
the left pattern in Figure~\ref{fig3}, hence, PMDS codes have
stronger requirements than optimal LRC codes.

Another family of codes with local and global properties specially
adapted to RAID type of 
architectures in which each storage device is an SSD, is given by the so
called Sector-Disk (SD) codes~\cite{pb}\cite{pbh}, which are closely
related to PMDS codes. These
codes can tolerate one or more device failures, and in addition, a
number of page failures. Like in the case of Figure~\ref{fig3}, we
illustrate the correction power of a (4,5;1,2) SD code in Figure~\ref{fig4}.

\begin{figure}
$$
\begin{array}{cc}
\begin{array}{|c|c|c|c|c|}
\hline
X&\phantom{X}&X&\phantom{X}&\phantom{X}\\
\hline
\phantom{X}&\phantom{X}&X&\phantom{X}&\phantom{X}\\
\hline
\phantom{X}&\phantom{X}&X&\phantom{X}&X\\
\hline
\phantom{X}&\phantom{X}&X&\phantom{X}&\phantom{X}\\
\hline
\end{array}
&
\begin{array}{|c|c|c|c|c|}
\hline
\phantom{X}&\phantom{X}&X&\phantom{X}&\phantom{X}\\
\hline
\phantom{X}&X&X&\phantom{X}&X\\
\hline
\phantom{X}&\phantom{X}&X&\phantom{X}&\phantom{X}\\
\hline
\phantom{X}&\phantom{X}&X&\phantom{X}&\phantom{X}\\
\hline
\end{array}
\end{array}
$$
\caption{\label{fig4}
Patterns that can be corrected by a (4,5;1,2) SD code}
\end{figure}

We can see that in both arrays in Figure~\ref{fig4}, the third device
(represented by the third column), has had a total failure. In
addition, two random symbol (page) failures have occurred: in the
array in the left in two different rows, while in the array on the
right in the same row, and both
situations are corrected when the code is an SD code. Certainly, a
PMDS code like the one depicted 
in Figure~\ref{fig3} can also do the job, but the converse is not
true: an SD code like the one depicted in Figure~\ref{fig4} cannot correct
in general patterns like the one depicted in the left of
Figure~\ref{fig3}. So, what is the 
advantage of using SD codes over PMDS codes? The idea is, given that
the requirements are less stringent, to use a smaller finite field
for SD codes than for PMDS codes. For example,
constructions of $(m,n;\ell,2)$ PMDS and SD codes are presented
in~\cite{bpsy}. In these constructions, the size $q$ of the finite
field satisfies $q\geq mn$ for the SD codes, while $q$ is
roughly larger than $2mn$ for PMDS codes (obtaining general efficient
constructions of PMDS and SD codes is still an open problem).

Given the considerations above, it is desirable to have LRC codes 
having a relatively small field size. Operations over a small field have
less complexity than over a larger field due to the smaller look-up
tables required. Specifically, we will use Integrated Interleaved
(II) codes~\cite{hapkt}\cite{tk}  
over $GF(q)$, where $q\geq \max\{m,n\}$, as $(m,n;\ell,g)$ LRC codes.
Certainly, this requires a tradeoff between minimum distance and
finite field size.

In general, an II code is not an optimal LRC code since its minimum
distance does not achieve the bound given by~(\ref{dist}).
However, we will see that in some cases the minimum distance is not
the crucial parameter in the performance of LRC codes, but the
average number of erasures to data loss. We will show that with
respect to this parameter, the versatility in the choice of
parameters of II codes allows them to often outperform optimal LRC codes.

As related work, we remark that STAIR codes~\cite{ll}, 
similarly to II codes, use fields of small size. STAIR codes assume
correlations in sector 
failures in order to make corrections. In this work, we do not assume
any correlations between sectors or pages.

We assume that each symbol is protected
by one local group, but let us mention work
considering multiple localities~\cite{tb}\cite{zy}.

Other related work consists of the so called Zigzag
codes~\cite{twb}, in which an $m\times n$ array keeps the MDS property
on columns and optimizes the minimum
number of updates in the presence of one (column) failure.

II codes~\cite{hapkt}\cite{tk} are strongly related to
Generalized Concatenated (GC) codes~\cite{bz}\cite{du}\cite{z}. In
fact, II codes were constructed with the goal of giving
an explicit implementation of GC codes that is convenient in
applications like magnetic recording. Related codes are the two-level
coding used in IBM magnetic recording products in the 80s~\cite{pa}
and its extensions~\cite{ah}. II and related codes were designed for
correction of errors. In this paper we exploit their implicit two dimensional
structure for use as $(m,n;\ell,g)$ LRC codes. To this end we need to
prove some properties that are tailored to our erasure model.

The paper is structured as follows: in Section~\ref{GC} we give the
definition of II codes and prove their main erasure-correcting
property. We derive the
minimum distance of the codes as a corollary of the main property
(a result given without proof in~\cite{tk}). In
Section~\ref{LRC}, we briefly discuss implementation in practice of
the codes and then we
give some performance comparisons with other LRC codes, like optimal
LRC codes and PMDS codes. Depending on the model and on the failure
statistics, we argue
that the minimum distance is not always the best parameter to measure the
performance of LRC codes. The average number of failures (in what
follows, failures and erasures are used interchangeably) to data
loss instead may be more important. These two parameters are certainly
completely correlated for MDS codes,
but we show that this is not the case for LRC codes. In particular, we show
that II codes, although they are codes over a much smaller field, often
outperform optimal LRC codes when the parameter considered is the
average number of failures to data loss. We end the paper by drawing some
conclusions.

\section{Integrated Interleaved (II) MDS Codes as LRC Codes}
\label{GC}
We assume that the II codes that we describe in this section are $m\times n$ array
codes with symbols in a finite field $GF(q)$ of characteristic 2,
i.e., $q\eq 2^b$. 
In fact, the codes can be described over any finite field of
characteristic $p$, $p$ a prime number, but we keep $p\eq 2$ for
simplicity and because it is the case more relevant in applications.
Reading the
symbols horizontally in a row-wise manner gives a code of length
$mn$.

\begin{defin}
\label{defII}
{\rm
Consider a set $\{\C_i\}$ of $t$ linear $[n,k_i,u_i+1]$ codes  over
$GF(q)$ such that
$\C_{t-1}\subset \C_{t-2}\subset \ldots \subset \C_0$ and
$1\leq u_0<u_1<\ldots <u_{t-1}\leq n-1$.
Let $\uu$ be the following vector of non-decreasing integers and length
$m\eq s_0+s_1+\cdots +s_{t-1}$, where $s_i\geq 1$ for $0\leq i\leq t-1$:

\begin{eqnarray}
\label{equu}
\uu &=&
\left(\overbrace{u_0,u_0,\ldots,u_0}^{s_0},\overbrace{u_1,u_1,\ldots,u_1}^{s_1},\ldots,
\overbrace{u_{t-1},u_{t-1},\ldots,u_{t-1}}^{s_{t-1}}\right).
\end{eqnarray}

Let $\hs_t\eq 0$ and $\hs_i\eq \sum_{j=i}^{t-1}s_j$ for $0\leq i\leq t-1$ (in
particular, $\hs_{t-1}\eq s_{t-1}$).

Consider the code $\C(n;\uu)$ consisting of
$m\times n$ arrays over $GF(q)$ such that, given
an array with rows $\uc_0,\uc_1,\ldots,\uc_{m-1}$, then $\uc_j\in\C_0$ for $0\leq
j\leq m-1$ and, if $\al$ is a primitive element in $GF(q)$,

\begin{eqnarray}
\label{IIgen}
\bigoplus_{j=0}^{m-1}\al^{rj}\uc_j&\in& \C_{t-i}\;\;{\rm for}\;\; 1\leq
i\leq t-1\;\; {\rm and}\;\; \hs_{t-i+1}\leq r\leq \hs_{t-i}-1.
\end{eqnarray}

Then we say that $\C(n;\uu)$ is a $t$-level II code.
\qed
}
\end{defin}

Notice that in Definition~\ref{defII} we have made no assumptions on
the size of the field $GF(q)$.
From now on we assume that the codes $\{\C_i\}$, $0\leq i\leq t-1$,
are MDS, i.e., they are $[n,n-u_i,u_i+1]$ codes (generally, we will
choose RS or extended RS codes as MDS codes, although some other
possibilities will
be discussed in Section~\ref{LRC}). Hence, we assume $q\geq n$.
In~(\ref{IIgen}) the
rows $\uc_j$, $0\leq j\leq m-1$, are multiplied
by powers of $\al$ constituting the parity-check matrix of an
$[m,m-\hs_1]\eq [m,s_0]$ RS code over $GF(q)$. We will require this
code to be MDS also, so, $q\geq m+1$, thus, in
Definition~\ref{defII}, from now on we have, $$q\geq\max\{m+1\,,\,n\}.$$

Let us illustrate the construction of $\C(n;\uu)$ with some
examples.

\begin{ex}
\label{ex11}
{\em
Assume that $\C(n;\uu)$ is a 1-level II code, i.e., $t\eq 1$ and $\uu\eq
\left(\overbrace{u_0,u_0,\ldots,u_0}^{m}\right)$. Then, according
to Definition~\ref{defII}, we have $m\times n$ arrays such that each
row $u_i$ is in the code $\C_0$, which is an $[n,n-u_0]$ MDS code,
i.e., it can correct up to $u_0$ erasures.

\qed
}
\end{ex}

\begin{ex}
\label{ex22}
{\em
Assume $t\eq 2$, i.e., $\uu\eq
\left(\overbrace{u_0,u_0,\ldots,u_0}^{s_0},\overbrace{u_1,u_1,\ldots,u_1}^{s_1}\right)$,
$s_0+s_1\eq m$
and $\C(n;\uu)$ is a 2-level II code. Code $\C_0$ is an $[n,n-u_0]$
MDS code and code $\C_1$ is an $[n,n-u_1]$ code, where $1\leq u_0\,<\,u_1\,<\,n$.

Consider an $m\times n$ array in $\C(n;\uu)$, where the rows are
given by $\uc_j$, $0\leq j\leq m-1$ and $\uc_j\in\C_0$.
Then, according to~(\ref{IIgen}),

\begin{eqnarray}
\label{IIgent2}
\bigoplus_{j=0}^{m-1}\al^{rj}\uc_j&\in& \C_1\;\;{\rm for}\;\; 0\leq r\leq s_{1}-1.
\end{eqnarray}

Codes of this type were presented in~\cite{hapkt}, while a parity-check matrix was given
in~\cite{hl}.

For example, consider $\C(n;(u_0,u_1,u_1,u_1))$ over $GF(q)$, then, according
to~(\ref{IIgent2}), given a $4\times n$ array with rows
$\uc_0,\uc_1,\uc_2,\uc_3$, each $\uc_j\in\C_0$ and

$$
\begin{array}{ccrcrcrcl}
\uc_0&\xor &\uc_1&\xor &\uc_2&\xor &\uc_3 &\in &\C_1\\
\uc_0&\xor &\al\uc_1&\xor &\al^2\uc_2&\xor &\al^3\uc_3 &\in &\C_1\\
\uc_0&\xor &\al^2\uc_1&\xor &\al^4\uc_2&\xor &\al^6\uc_3 &\in &\C_1.\\
\end{array}
$$

\qed
}
\end{ex}

\begin{ex}
\label{ex33}
{\em
Assume $t\eq 3$, i.e., $\uu\eq
\left(\overbrace{u_0,u_0,\ldots,u_0}^{s_0},\overbrace{u_1,u_1,\ldots,u_1}^{s_1},
\overbrace{u_2,u_2,\ldots,u_2}^{s_2}\right)$,\\
$s_0+s_1+s_2\eq m$
and $\C(n;\uu)$ is a 3-level II code. Code $\C_0$ is an $[n,n-u_0]$
MDS code, code $\C_1$ is an $[n,n-u_1]$ code and code $\C_2$ is an
$[n,n-u_2]$ code, where $1\leq u_0\,<\,u_1\,<\,u_2\,<\,n$.

Consider an $m\times n$ array in $\C(n;\uu)$, where the rows are
given by $\uc_j$, $0\leq j\leq m-1$ and $\uc_j\in\C_0$.
Then, according to~(\ref{IIgen}),

\begin{eqnarray}
\label{IIgent31}
\bigoplus_{j=0}^{m-1}\al^{rj}\uc_j&\in& \C_2\;\;{\rm for}\;\; 0\leq
r\leq s_{2}-1\\
\label{IIgent32}
\bigoplus_{j=0}^{m-1}\al^{rj}\uc_j&\in& \C_1\;\;{\rm for}\;\; s_2\leq
r\leq s_{1}+s_2-1
\end{eqnarray}

For example, consider $\C(n;(u_0,u_1,u_1,u_2))$ over $GF(q)$, then,
given a $4\times 5$ array with rows
$\uc_0,\uc_1,\uc_2,\uc_3$, each $\uc_j\in\C_0$ and, since $s_1\eq 2$
and $s_2\eq 1$, (\ref{IIgent31}) and~(\ref{IIgent32}) give

$$
\begin{array}{ccrcrcrcl}
\uc_0&\xor &\uc_1&\xor &\uc_2&\xor &\uc_3 &\in &\C_2\\
\uc_0&\xor &\al\uc_1&\xor &\al^2\uc_2&\xor &\al^3\uc_3 &\in &\C_1\\
\uc_0&\xor &\al^2\uc_1&\xor &\al^4\uc_2&\xor &\al^6\uc_3 &\in &\C_1.\\
\end{array}
$$

\qed
}
\end{ex}

We have seen that in~(\ref{IIgen}) the rows $\uc_j$, $0\leq j\leq
m-1$, are multiplied
by powers of $\al$ constituting the parity-check matrix of an
$[m,m-\hs_1,\hs_1+1]$ RS code over $GF(q)$. In fact, we can also
use the parity-check matrix of a (shortened)
extended $[m,m-\hs_1,\hs_1+1]$ RS code over $GF(q)$. In this case, it
suffices with taking $q\geq m$, hence, $$q\geq\max\{m\,,\,n\}.$$
In this case, instead of~(\ref{IIgen}) we may have

\begin{eqnarray}
\label{IIgenext0}
\bigoplus_{j=0}^{m-1}\uc_j\phantom{\al^{rj}}&\in& \C_{t-1}\\
\label{IIgenext1}
\bigoplus_{j=0}^{m-2}\al^{rj}\uc_j&\in& \C_{t-1}\;\;{\rm for}
\;\; 1\leq r\leq s_{t-1}-1\\
\label{IIgenext}
\bigoplus_{j=0}^{m-2}\al^{rj}\uc_j&\in& \C_{t-i}\;\;{\rm for}\;\; 2\leq
i\leq t-1\;\; {\rm and}\;\; \hs_{t-i+1}\leq r\leq \hs_{t-i}-1.
\end{eqnarray}

The advantage of using~(\ref{IIgenext0}), (\ref{IIgenext1})
and~(\ref{IIgenext}) instead of~(\ref{IIgen}) is that when $m$ is a
power of 2, a smaller field is required.
We illustrate this situation in the next example.

\begin{ex}
\label{ex333}
{\em
Consider a $4\times 4$ array in the 3-level II code $\C(4;(1,2,2,3))$
over $GF(4)$, where 
the rows are given by $\uc_j$, $0\leq j\leq 3$,
$\C_i$ is a $[4,4-i-1,i+2]$ extended RS code over $GF(4)$ for $0\leq
i\leq 2$ and
$\uc_j\in\C_0$.
Then, according to~(\ref{IIgenext0}), (\ref{IIgenext1})
and~(\ref{IIgenext}), if $\al$ is primitive in $GF(4)$,

$$
\begin{array}{ccrcrcrcl}
\uc_0&\xor &\uc_1&\xor &\uc_2&\xor &\uc_3 &\in &\C_2\\
\uc_0&\xor &\al\uc_1&\xor &\al^2\uc_2& & &\in &\C_1\\
\uc_0&\xor &\al^2\uc_1&\xor &\al\,\uc_2& & &\in &\C_1.\\
\end{array}
$$

If we used~(\ref{IIgen}) instead of~(\ref{IIgenext0}), (\ref{IIgenext1})
and~(\ref{IIgenext}), we would require a field of size $q$, with
$q\geq \max\{5,4\}\eq 5$, so the smallest field we could use would be
$GF(8)$ instead of $GF(4)$.

\qed
}
\end{ex}

Next we give the main property of $t$-level II codes.

\begin{theo}
\label{theo1}
{\em
Consider the $t$-level II code $\C(n;\uu)$ over $GF(q)$ as given by
Definition~\ref{defII} with either condition~(\ref{IIgen}) or
conditions~(\ref{IIgenext0}), (\ref{IIgenext1})
and~(\ref{IIgenext}), where the component codes are MDS and
$q\,\geq\,\max\{m+1,n\}$ when~(\ref{IIgen}) holds, and
$q\,\geq\,\max\{m,n\}$ when~(\ref{IIgenext0}), (\ref{IIgenext1})
and~(\ref{IIgenext}) hold.
Then, $\C(n;\uu)$ can correct up to $u_0$ erasures in any row, and up
to $u_i$ erasures in any $s_i$ rows, $1\leq i\leq t-1$, of an $m\times n$ array
corresponding to a codeword in $\C(n;\uu)$.
}
\end{theo}

Before formally proving Theorem~\ref{theo1}, we illustrate it with an
example.

\begin{ex}
\label{ex44}
{\em
Consider again code $\C(n;(u_0,u_1,u_1,u_2))$ over $GF(q)$ as depicted
in Example~\ref{ex33}. According to Theorem~\ref{theo1}, the code can
correct any row with up to $u_0$ erasures, two rows with up to $u_1$
erasures each and one
row with up to $u_2$ erasures, where $u_0<u_1<u_2$. Specifically, assume that
we have an array with rows $\uc_0$, $\uc_1$, $\uc_2$ and $\uc_3$ such
that row $\uc_2$ has $u_0$ erasures,
rows $\uc_0$ and $\uc_3$ have $u_1$ erasures each and row $\uc_1$ has $u_2$ erasures.

Since each row is in $\C_0$, we first correct the $u_0$ erasures in
$\uc_2$. Then, according to~(\ref{IIgent31}) and~(\ref{IIgent32}),
reordering the rows $\uc_i$ in decreasing number of erasures, we have

$$
\begin{array}{rrrrrrrcl}
\uc_1&\xor&\uc_0&\xor&\uc_3&\xor&\uc_2 &\;\in\; &\C_2\\
\al\uc_1&\xor &\uc_0&\xor&\al^3\uc_3&\xor&\al^2\uc_2 &\in &\C_1\\
\al^2\uc_1&\xor&\uc_0&\xor&\al^6\uc_3&\xor&\al^4\uc_2 &\in &\C_1.
\end{array}
$$

Since the coefficients form a Vandermonde matrix, we can triangulate
the system above, and since $\C_2\subset \C_1$, we obtain

$$
\begin{array}{rrrrrrrcl}
\uc_1&\xor&\uc_0&\xor&\uc_3&\xor&\uc_2 &\;\in\; &\C_2\\
&&\uc_0&\xor&\ga_{1,3}\uc_3&\xor&\ga_{1,2}\uc_2 &\in &\C_1\\
&&&&\uc_3&\xor&\ga_{2,2}\uc_2 &\in &\C_1,
\end{array}
$$
the coefficients $\ga_{i,j}$ obtained as a result of the
triangulation. In particular, since $\uc_3$ has $u_1$ erasures, also
$\uc_3\xor\ga_{2,2}\uc_2$ has $u_1$ erasures. Since
$\uc_3\xor\ga_{2,2}\uc_2\in \C_1$, the erasures can be corrected. Once
the erasures are corrected, $\uc_3$ is obtained by XORing this
corrected vector with $\ga_{2,2}\uc_2$. Then vector
$\uc_0\xor\ga_{1,3}\uc_3\xor\ga_{1,2}\uc_2$ has $u_1$ erasures, which
are corrected, and $\uc_0$ is obtained by XORing the corrected vector
with $\ga_{1,3}\uc_3\xor\ga_{1,2}\uc_2$. Finally,
$\uc_1\xor\uc_0\xor\uc_3\xor\uc_2$ has $u_2$ erasures, but this vector
is in $\C_2$, so the erasures are corrected and $\uc_1$ is obtained
by XORing the corrected vector with $\uc_0\xor\uc_3\xor\uc_2$.

The proof of Theorem~\ref{theo1} generalizes this procedure, which
also provides a decoding algorithm.
\qed
}
\end{ex}

\noindent
{\bf Proof of Theorem~\ref{theo1}:}
We will prove the result by assuming condition~(\ref{IIgen}), the
proof for
conditions~(\ref{IIgenext0}), (\ref{IIgenext1})
and~(\ref{IIgenext}) being completely analogous.

Assume that there are at least $s_0$ rows with up to $u_0$ erasures each.
Since each row is in $\C_0$, the erasures in these
rows can be corrected.  So, without loss of generality, assume that
there are no rows with at most $u_0$ erasures, and
for each $i$, $1\leq i\leq t-1$, there are $s'_i$ rows with
more than $u_{i-1}$ erasures and at most $u_i$ erasures,
where $0\leq s'_i\leq s_i$.  We
proceed by induction on the total number of rows with erasures $s$.

Assume first that $s\eq 1$. Then we have exactly one row $\uc_v$,
$0\leq v\leq m-1$, with at most $u_{t-1}$ erasures. In particular,
taking $i\eq 1$ and $r\eq 0$ in~(\ref{IIgen}),
$$\uc\;\eq\; \bigoplus_{j=0}^{m-1}\uc_j\in \C_{t-1},$$ so
$\uc$ has at most $u_{t-1}$ erasures, which can be corrected. Then,
once $\uc$ is corrected, we obtain $\uc_v$ as
$$\uc_v\eq \uc\,\xor\, \bigoplus_{j=0\atop j\neq v}^{m-1}\uc_j.$$

Assume next that $s\,>\,1$.
Partition the set $\{0,1,\ldots,m-1\}$ into $t$
disjoint sets $S_i$, such that set
$S_0$ consists of the locations of rows with no erasures, and for
$1\leq i\leq t-1$,
set $S_{i}$ consists of the locations of rows
with more than $u_{i-1}$ erasures and at most
$u_i$ erasures
(in Example~\ref{ex44}, $s\eq 3$, $S_0\eq\{2\}$, $S_1\eq\{0,3\}$ and
$S_2\eq\{1\}$; notice also that $|S_i|\eq s'_i$ and that $S_i$ may be empty).

Rearranging~(\ref{IIgen})
following the order given by $S_{t-1},S_{t-2},\ldots,S_0$, we obtain

\begin{eqnarray}
\label{IIgenper}
\bigoplus_{v=1}^{t}\bigoplus_{u\in S_{t-v}}\al^{ru}\uc_u&\in& \C_{t-i}\;\;{\rm for}\;\; 1\leq
i\leq t-1\;\; {\rm and}\;\; \hs_{t-i+1}\leq r\leq \hs_{t-i}-1.
\end{eqnarray}

Let $$w_0\;\eq\; \min\{w\,:\,S_{w}\neq \emptyset\;\;{\rm
for}\;\;w\geq 1\}.$$
In particular, since $\C_{t-1}\subset \C_{t-2}\subset \ldots \subset
\C_{w_0}$ and taking the first $s$ rows in~(\ref{IIgenper}), we obtain

\begin{eqnarray}
\label{IIgenperw}
\bigoplus_{v=1}^{t}\bigoplus_{u\in S_{t-v}}\al^{ru}\uc_u&\in&
\C_{w_0}\;\;{\rm for}\;\; 0\leq r\leq s-1.
\end{eqnarray}

Let $r_L$ be the last element in set $S_{w_0}$
(each set $S_u$ can be ordered in increasing order; if so, in
Example~\ref{ex44}, $w_0\eq 1$ and $r_L\eq 3$).
Since the matrix of coefficients $\al^{ru}$ in~(\ref{IIgenperw}) is
Vandermonde, in particular, it
can be triangulated. The last row of this triangulation is

\begin{eqnarray*}
\uc_{r_L}\xor \bigoplus_{u\in S_{0}}\ga_{u}\uc_u&\in& \C_{w_0},
\end{eqnarray*}
where the coefficients $\ga_u$ are obtained as a consequence of the
triangulation. In particular, $\uc_{r_L}\xor
\bigoplus_{u\in S_{0}}\ga_{u}\uc_u$ is in $\C_{w_0}$
since it is a linear combination of elements in $\C_{w_0}$.
Since $\uc_{r_L}$
has $u'_{w_0}$ erasures, where $u_{w_0-1}<u'_{w_0}\leq u_{w_0}$,
also $\uc_{r_L}\xor
\bigoplus_{u\in S_{0}}\ga_{u}\uc_u$ has $u'_{w_0}$
erasures, which can be
corrected. Once the erasures are corrected, $\uc_{r_L}$ is obtained by
XORing the corrected vector with
$\bigoplus_{u\in S_{0}}\ga_{u}\uc_u$. This leaves us with $s-1$ rows with
erasures and the result follows by induction.
\qed

As is the case in general in erasure decoding, the encoding is a
special case of
the decoding. For example, in a $t$-level II code $\C(n;\uu)$ code as
given by
Definition~\ref{defII}, we may dedicate to parity the last $u_0$ symbols
of the first $s_0$ rows, the last $u_1$ symbols of the following
$s_1$ rows, and so on, until the last $u_{t-1}$ symbols of the last
$s_{t-1}$ rows. If in Example~\ref{ex44} we take $u_0\eq
1$, $u_1\eq 2$ and $u_3\eq 3$, the allocation of data and parity
would be as depicted in Figure~\ref{fig5}.

\begin{figure}
$$
\begin{array}{|c|c|c|c|c|}
\hline
{\bf D}&{\bf D}&{\bf D}&{\bf D}&{\bf L}\\
\hline
{\bf D}&{\bf D}&{\bf D}&{\bf G}&{\bf L}\\
\hline
{\bf D}&{\bf D}&{\bf D}&{\bf G}&{\bf L}\\
\hline
{\bf D}&{\bf D}&{\bf G}&{\bf G}&{\bf L}\\
\hline
\end{array}
$$
\caption{\label{fig5}
Allocation of data and parity for code $\C(5;(1,2,2,3))$.}
\end{figure}

This also gives us the dimension of code $\C(n;\uu)$.

\begin{cor}
\label{cor0}
{\em
Consider the $t$-level II code $\C(n;\uu)$ of
Theorem~\ref{theo1}. Then, $\C(n;\uu)$ is an
$[mn,mn-\sum_{i=0}^{m-1}s_iu_i]$ code.
}
\end{cor}

The following result was given without proof in~\cite{tk}:

\begin{cor}
\label{cor1}
{\em
Consider the $t$-level II code $\C(n;\uu)$ of
Theorem~\ref{theo1}. Then, if $\hs_t\eq 0$ and $\hs_i\eq
\sum_{j=i}^t s_j$ for $0\leq i\leq t-1$, the minimum
distance of $\C(n;\uu)$ is given by

\begin{eqnarray*}
d&=&\min\left\{\left(\hs_{i+1}+1\right)\left(u_i+1\right)\;,\;0\leq
i\leq t-1\right\}.
\end{eqnarray*}
}
\end{cor}

\noindent\pf 
For each $i$ such that $0\leq i\leq t-1$, consider an array
in which $\hs_{i+1}$ rows have $u_i+1$ erasures each, one row
has $u_i$ erasures, and all the other entries are zero 
(when $i\eq t-1$, this means that there
is a row with $u_{t-1}$ erasures and all the other entries are
zero). By Theorem~\ref{theo1}, such an array would be corrected by
the code $\C(n;\uu)$ as the zero codeword, thus

\begin{eqnarray*}
d&\geq &\min\left\{\left(\hs_{i+1}+1\right)\left(u_i+1\right)\;,\;0\leq
i\leq t-1\right\}.
\end{eqnarray*}

In order to show equality, we need to prove that if $w$ satisfies

\begin{eqnarray*}
\left(\hs_{w+1}+1\right)\left(u_w+1\right)&\eq &\min\left\{\left(\hs_{i+1}+1\right)\left(u_i+1\right)\;,\;0\leq
i\leq t-1\right\}, 
\end{eqnarray*}
then there is a codeword in $\C(n;\uu)$ of weight $\left(\hs_{w+1}+1\right)\left(u_w+1\right)$.
We will proceed by assuming condition~(\ref{IIgen}), the proof for
conditions~(\ref{IIgenext0}), (\ref{IIgenext1})
and~(\ref{IIgenext}) being completely analogous.

Let $\uu$ be a codeword of
weight $u_w+1$ in $\C_w$.
Let $\uv$ be a
codeword of weight $\hs_{w+1}+1$ in the $[\hs_{w+1}+1,1,\hs_{w+1}+1]$ RS code whose
parity-check matrix is given by
$$
\left(
\begin{array}{ccccc}
1&1&1&\ldots &1\\
1&\al &\al^2& \ldots &\al^{\hs_{w+1}}\\
1&\al^2 &\al^4& \ldots &\al^{2\hs_{w+1}}\\
\vdots &\vdots &\vdots &\ddots &\vdots \\
1&\al^{\hs_{w+1}-1} &\al^{2(\hs_{w+1}-1)}& \ldots &\al^{(\hs_{w+1}-1)\hs_{w+1}}\\
\end{array}
\right)
$$

Explicitly, let $\uv\eq (v_0,v_1,\ldots,v_{\hs_{w+1}})$. In
particular,

\begin{eqnarray}
\label{ref}
\bigoplus_{j=0}^{\hs_{w+1}}\al^{rj}v_j&\eq &0\;\;{\rm for}\;\;0\leq
r\leq \hs_{w+1}-1.
\end{eqnarray}

Consider the $m\times n$ array of weight
$\left(\hs_{w+1}+1\right)\left(u_w+1\right)$ whose rows are:
\begin{eqnarray*}
\left(v_0\,\uu\;,\;v_1\,\uu\;,\;\ldots\;,\;v_{\hs_{w+1}}\uu\;,\;
\overbrace{\uzero_n\;,\;\uzero_n\;,\;\ldots\;,\;\uzero_n}^{m-\hs_{w+1}-1}\right),
\end{eqnarray*}
where $\uzero_n$ denotes the zero
vector of length $n$. We will show that this array is in $\C(n;\uu)$.
According to~(\ref{IIgen}), we have to show that

\begin{eqnarray}
\label{IIgenw}
\bigoplus_{j=0}^{\hs_{w+1}}\al^{rj}\left(v_j\,\uu\right)
&\in& \C_{t-i}\;\;{\rm for}\;\; 1\leq
i\leq t-1\;\; {\rm and}\;\; \hs_{t-i+1}\leq r\leq \hs_{t-i}-1.
\end{eqnarray}

Assume first that $t-i\leq w\leq t-1$. Since $\C_w\subseteq
\C_{t-i}$, in particular, $\uu\in\C_{t-i}$ 
and hence
$\bigoplus_{j=0}^{\hs_{w+1}}\al^{rj}\left(v_j\,\uu\right)\in\C_{t-i}$,
so~(\ref{IIgenw}) follows.

Assume next that $0\leq w\leq t-i-1$. Then, $\hs_{w+1}\geq \hs_{t-i}$,
and, for $0\leq r\leq \hs_{t-i}-1$, by~(\ref{ref}),
\begin{eqnarray*}
\bigoplus_{j=0}^{\hs_{w+1}}\al^{rj}\left(v_j\,\uu\right)\;\;\eq\;\;
\left(\bigoplus_{j=0}^{\hs_{w+1}}\al^{rj}v_j\right)\,\uu&\eq
&\uzero_n.\end{eqnarray*}
Since $\uzero_n\in\C_{t-i}$, (\ref{IIgenw}) follows also in this case.
\qed

\begin{ex}
\label{ex55}
{\em
Consider again code $\C(n;(u_0,u_1,u_1,u_2))$ over $GF(q)$ as
in Examples~\ref{ex33} and~\ref{ex44}. Corollary~\ref{cor1} states that the minimum
distance of $\C(n;(u_0,u_1,u_1,u_2))$ is given by

\begin{eqnarray*}
d&=&\min\left\{(4)(u_0+1)\;,\;(2)(u_1+1)\;,\;u_2+1
\right\}.
\end{eqnarray*}

In the case depicted in Figure~\ref{fig5}, $u_0\eq 1$, $u_1\eq 2$ and
$u_2\eq 3$,  so

\begin{eqnarray*}
d&=&\min\left\{(4)(2)\;,\;(2)(3)\;,\;4
\right\}\;\;=\;\;4.
\end{eqnarray*}

\qed
}
\end{ex}

Consider next a 2-level II code $\C(n;\uu)$ such that

\begin{eqnarray}
\label{equ1}
\uu &=&
\left(\overbrace{u_0,u_0,\ldots,u_0}^{s_0},u_1\right).
\end{eqnarray}

According to Corollary~\ref{cor1}, since $s_1\eq 1$, the minimum
distance of $\C(n;\uu)$ is given by

\begin{eqnarray*}
d&=&\min\left\{u_1+1\;,\;2(u_0+1)\right\}.
\end{eqnarray*}

Notice that this code has $\ell\eq u_0$ local parities per row and
$g\eq u_1-u_0$
global parities. Thus,
according to bound~(\ref{dist1}), $d\leq u_0+(u_1-u_0)+1\eq u_1+1$.
Therefore, $d$ achieves bound~(\ref{dist1})  (i.e., the code is an
optimal LRC code) if and only if $u_1+1\leq
2(u_0+1)$, if and only if $u_1\leq 2u_0+1$. Let us state this
observation as a corollary (this result was also given in~\cite{b}).

\begin{cor}
\label{cor2}
{\em
Consider the $2$-level II code $\C(n;\uu)$ with $\uu$ given
by~(\ref{equ1}) and $u_1\leq 2u_0+1$. Then, $\C(n;\uu)$ is an optimal
LRC code.
}
\end{cor}

Corollary~\ref{cor2} can be interpreted as, given parameters $m$,
$n$, $\ell$ and $g$ with $\ell +g\,<\,n$ and $g\leq \ell +1$, there exists
an optimal $(m,n;\ell,g)$ LRC code over a field of 
size $q\geq\max\{m,n\}$. The general constructions of optimal LRC
codes~\cite{tb} require a field of size $q\geq mn$.

\section{Implementation and Performance}
\label{LRC}
Consider a $t$-level II code $\C(n;\uu)$ as given by Definition~\ref{defII}.
The proof of Theorem~\ref{theo1} provides a decoding algorithm for
the erasures within the correcting capability of the code. As stated
in Theorem~\ref{theo1}, the first step involves correcting those rows
having
up to $u_0$ erasures each. The second step involves a triangulation
after which a row with up to $u_1$ erasures is corrected. The
triangulation is done only once since once this row is corrected,
assuming the triangulated matrix has $s$ rows, we proceed with the
first $s-1$ rows of this triangulated matrix and repeat the process.
For more implementation details, 
see~\cite{bh}.

The decoding algorithm is tailored for erasures, but it can be
adapted for errors as well. For decoding algorithms correcting errors
using II codes, see~\cite{cw}\cite{tk}\cite{w}. 

A convenient implementation of II codes may be done by
using the ring of polynomials modulo $1+x+x^2+\cdots
+x^{p-1}$, where $p$ is a prime and $p\geq \min\{m,n\}$,
instead of a field. This ring
allows for using symbols of large size and avoiding look-up tables:
multiplications by powers of $\alpha$ are 
basically rotations. The individual MDS codes
$\C_i$ may be, for instance,
Blaum-Roth (BR) codes~\cite{br} or extended EVENODD codes~\cite{bbbmv}.
For reasons of space, we omit the details.

Next we establish performance comparisons between $t$-level II codes
and other types of LRC codes, such as optimal LRC and PMDS codes. Given
an $m\times n$ array, as stated, the main advantage of II codes is that
the size of the field required is much smaller: in effect,
for optimal LRC codes the size of the field is in general $q\geq mn$,
while for II codes it is $q\geq \max\{m,n\}$.
We will argue next that when the parameters are carefully chosen, the
performance of II codes is competitive with the one of optimal LRC
codes.

Optimal LRC codes have in general better minimum distance
than II codes, the exception being given by codes satisfying the
conditions of Corollary~\ref{cor2}. However, the minimum distance,
although still a very
important parameter, is not always the most important one reflecting
the performance of the code. We will see this in two ways.

Firstly, assume that the probability of a single erasure is $P$. The
probability of data loss is usually dominated by
the first term of the weight distribution, i.e., the number of
codewords of weight $d$. However, if, say, the number of codewords of
weight $d+1$ is much larger than the number of codewords of weight
$d$, and the probability $P$ is not too small (for example, the
failure rate of SSDs is reported to be 1.5\% in a year~\cite{me}, or
$P\eq .015$), the
second term in the probability of data loss may be larger than the
first. Let us illustrate this situation with a concrete case.

Consider two $(m,n;1,3)$ LRC codes with $n\,>\,4$.
One of them is an optimal LRC code, which by bound~(\ref{dist1}) has
minimum distance $d\eq 5$. The second one is a 3-level II code
$\C(n;(\overbrace{1,1,\ldots,1}^{m-2},2,3))$, which by
Corollary~\ref{cor1} has minimum distance $d\eq 4$. We will compute
the probability of data loss under erasures for both of them and study under which
parameters one probability is larger than the other one.

Consider first the patterns of erasures that can be corrected by
$\C(n;(\overbrace{1,1,\ldots,1}^{m-2},2,3))$ but not by the
optimal LRC code. Since the later has minimum distance $d\eq 5$, by
Theorem~\ref{theo1}, this will occur only if at least 5 erasures have
occurred, a row has exactly two erasures, another row exactly three
erasures, and the remaining erasures are in one row each. It is not
hard to see that the probability of this occurring is

\begin{eqnarray}
\label{pLRC}
P_{\rm OLRC} &=& m(m-1){n\choose 2}{n\choose
3}P^5(1-P)^{mn-5}\sum_{i=0}^{m-2}{m-2\choose i}\left({nP\over 1-P}\right)^i\,.
\end{eqnarray}

Similarly, the cases that can be corrected by the optimal LRC code
but not by \\$\C(n;(\overbrace{1,1,\ldots,1}^{m-2},2,3)$ occur when one
row has exactly 4 erasures and the remaining rows with erasures have
at most one erasure. The probability of this occurring is

\begin{eqnarray}
\label{pII}
P_{\rm II} &=& m{n\choose 4}P^4(1-P)^{mn-4}\sum_{i=0}^{m-1}{m-1\choose i}\left({nP\over 1-P}\right)^i\,.
\end{eqnarray}

Consider the quotient

\begin{eqnarray}
\label{quot}
P_{\rm OLRC}/P_{\rm II}&=&
\left({2(m-1)n(n-1)\over n-3}\right)\left({P\over 1-P}\right)\left({\sum_{i=0}^{m-2}{m-2\choose
i}\left({nP\over 1-P}\right)^i\over\sum_{i=0}^{m-1}{m-1\choose i}\left({nP\over 1-P}\right)^i}\right)\,.
\end{eqnarray}

Whenever $P_{\rm OLRC}/P_{\rm II}\,>\,1$, the optimal $(m,n;1,3)$ LRC
code has a higher probability of data loss than the 3-level II code
$\C(n;(\overbrace{1,1,\ldots,1}^{m-2},2,3)$. Table~\ref{t0} gives
values of some different parameters $m$, $n$
and $P$ for which the quotient is (very close to) 1. For these values, both
$(m,n;1,3)$ LRC codes have roughly the same probability of data loss. For $P$
above the value given in Table~\ref{t0}, the optimal $(m,n;1,3)$ LRC
code has more probability of data loss than the 3-level II code
$\C(n;(\overbrace{1,1,\ldots,1}^{m-2},2,3)$. The probability $P$ is
multiplied by 100 so it is given as a percentage failure rate in Table~\ref{t0}.

\begin{table}
\begin{center}
\begin{tabular}{|c|c||c|}
\hline
$m$&$n$&$100P$\%\\
\hline\hline
8& 5 &.73\%\\
\hline
16& 5 &.34\%\\
\hline
8& 8 &.67\%\\
\hline
16& 8 &.31\%\\
\hline
8& 12 &.52\%\\
\hline
16& 12 &.24\%\\
\hline
32& 12 &.12\%\\
\hline
\end{tabular}
\end{center}
\caption{\label{t0}
Some parameters giving $P_{\rm OLRC}\approx P_{\rm II}$
for $(m,n;1,3)$ LRC codes.}
\end{table}

If SSDs are used and we assume that the annual
failure rate is around 1.5\% as suggested in~\cite{me}, then in all
cases of Table~\ref{t0} the 3-level II code
$\C(n;(\overbrace{1,1,\ldots,1}^{m-2},2,3)$ has superior performance.

Next we analyze another parameter for the performance of an
$(m,n;\ell,g)$ LRC code: the average number
of erasures that cause an uncorrectable pattern (and hence, data
loss), that we denote by
$Av_{\rm fail}$. This parameter is closely related to the Mean Time
to Data Loss (MTTDL) parameter, but we do not explore the connection
here. We argue that $Av_{\rm fail}$ may be more important
than the minimum distance in some applications.

In effect, assume that erasures occur at random in an $m\times n$
array. The model may correspond to a system of storage devices where
failures are being tracked. The idea is to allow for failures for as
long as possible, before requesting for maintenance. For example,
maintenance may be requested when the system is two failures away
from an uncorrectable pattern which would cause data loss (calling
for maintenance while being only one
failure away from data loss may be too risky). Since
maintenance is expensive, it is desirable
to delay it as much as possible. Let us point out that
similar models were considered for computer memories protected
against single errors~\cite{bgm}\cite{gm}\cite{gm2}. These references
also explore the connection between $Av_{\rm fail}$ and MTTDL.

The next simple example illustrates the concept of average number
of erasures causing an uncorrectable pattern.

\begin{ex}
\label{ex5}
{\em
Consider an $(m,n;1,0)$ LRC code, that is, an $m\times n$ array
with one local parity per row and no global parities (this
corresponds to a RAID~5 scheme). We will have data loss when two
erasures in the same row have occurred. So, what is the average
number of erasures until we have data loss? One way to do this is by
running a Montecarlo simulation and averaging over a large number of
trials. If we proceed like this, we find out, for example, that when
$m\eq 365$, then $Av_{\rm fail}\approx 24.6$. The reader may
recognize this number as the birthday surprise number:
assuming that people start arriving at random, how many people arrive
on average until two of them share the same birthday? There are exact
formulae to compute the birthday surprise number~\cite{gm2}\cite{kn}.
For example, in a planet with $m$ days, the birthday surprise number,
which is equivalent to our problem for the $(m,n;1,0)$ LRC code, is
given by

\begin{eqnarray*}
Av_{\rm fail}&=&m\int_0^{\infty}e^{-mx}(1+x)^m\,dx.
\end{eqnarray*}
It is possible to obtain formulae like the above one for more
complicated cases, but that is beyond the scope of this paper. In any
case, Montecarlo simulations give good approximations.
\qed
}
\end{ex}

\begin{table}
\begin{center}
\begin{tabular}{|c|c|c|c|c|c|c|}
\hline
Parameters&Code & $d_{\rm min}$ & $Av_{\rm fail}$\\
\hline\hline
[80,61]& [80,61] MDS &20 &20 \\
& (16,5;1,3) PMDS &5 &12.7 \\
&Optimal (16,5;1,3) LRC &5 &10.8 \\
&$\C(5;(\overbrace{1,1,\ldots 1}^{14},2,3)$ 3-level II &4 &11.6 \\
\hline
[80,60] & [80,60] MDS &21 &21 \\
& (16,5;1,4) PMDS &7 &14.5 \\
&Optimal (16,5;1,4) LRC &7 &13 \\
&$\C(5;\overbrace{1,1,\ldots 1}^{13},2,2,3)$ 3-level II &4 &13.5 \\
\hline
[80,59]& [80,59] MDS &22 &22 \\
& (16,5;1,5) PMDS &8 &16.2 \\
&Optimal (16,5;1,5) LRC &8 &14 \\
&$\C(5;\overbrace{1,1,\ldots 1}^{12},2,2,2,3)$ 3-level II &4 &15 \\
\hline
[80,58]& [80,58] MDS &23 &23 \\
& (16,5;1,6) PMDS &9 &17.7 \\
&Optimal (16,5;1,6) LRC &9 &15 \\
&$\C(5;\overbrace{1,1,\ldots 1}^{11},2,2,2,2,3)$ 3-level II &4 &16 \\
\hline
[80,57]& [80,57] MDS &24 &24 \\
& (16,5;1,7) PMDS &10 &19.2 \\
&Optimal (16,5;1,7) LRC &10 &16 \\
&$\C(5;\overbrace{1,1,\ldots 1}^{11},2,2,2,3,3)$ 3-level II &4 &17.1 \\
\hline
[80,56]& [80,56] MDS &25 &25 \\
& (16,5;1,8) PMDS &12 &20.6 \\
&Optimal (16,5;1,8) LRC &12 &18 \\
&$\C(5;\overbrace{1,1,\ldots 1}^{11},2,2,2,3,4)$ 4-level II &5 &18.5 \\
\hline
\end{tabular}
\end{center}
\caption{\label{t1}
Some codes corresponding to $16\times 5$ arrays}
\end{table}

\begin{table}
\begin{center}
\begin{tabular}{|c|c|c|c|c|c|}
\hline
Parameters&Code & $d_{\rm min}$ & $Av_{\rm fail}$\\
\hline\hline
[128,92]& [128,92] MDS &37 &37 \\
&(16,8;2,4) PMDS &7 &25.1 \\
&Optimal (16,8;2,4) LRC &7 &20.4 \\
&$\C(8;\overbrace{2,2,\ldots 2}^{13},3,3,4)$ 3-level II  &5 &23.8 \\
\hline
[128,91]& [128,91] MDS &38 &38 \\
&(16,8;2,5) PMDS &8 &27.8 \\
&Optimal (16,8;2,5) LRC &8 &21.8 \\
&$\C(8;\overbrace{2,2,\ldots 2}^{12},3,3,3,4)$ 3-level II  &5 &25 \\
\hline
[128,90]& [128,90] MDS &39 &39 \\
&(16,8;2,6) PMDS &11 &29.1 \\
&Optimal (16,8;2,6) LRC &11 &24.9 \\
&$\C(8;\overbrace{2,2,\ldots 2}^{12},3,3,4,4)$ 3-level II  &5 &26.3 \\
\hline
[128,89] & [128,89] MDS &40 &40 \\
&(16,8;2,7) PMDS &12 &30.1 \\
&Optimal (16,8;2,7) LRC &12 &25.7 \\
&$\C(8;\overbrace{2,2,\ldots 2}^{12},3,3,4,5)$ 4-level II  &6 &27.5 \\
\hline
[128,84]& [128,84] MDS &45 &45 \\
&(16,8;2,12) PMDS &19 &38.8 \\
&Optimal (16,8;2,12) LRC &19 &32.2 \\
&$\C(8;\overbrace{2,2,\ldots 2}^{10},3,3,3,4,5,6)$ 5-level II  &7 &34.7 \\
\hline
\end{tabular}
\end{center}
\caption{\label{t2}
Some codes corresponding to $16\times 8$ arrays}
\end{table}

Notice that, if a code is MDS, the minimum distance $d$ and $Av_{\rm
fail}$ are completely correlated. In fact, data loss for an MDS
code occurs each time there are $d$ erasures and not before, thus,
$Av_{\rm fail}\eq d$. But this property is lost for LRC codes, and we
will show next that in many cases, LRC codes with better minimum
distance $d$ than others have however worse $Av_{\rm fail}$.

Consider in general $(m,n;\ell,g)$ LRC codes. In terms of $Av_{\rm
fail}$, the best we can do is a PMDS code, since it can correct all
possible patterns under the locality and number of global parities
restrictions. So, $Av_{\rm fail}$ for an $(m,n;\ell,g)$ PMDS code
provides an upper bound.

Next consider
$t$-level II codes $\C(n;\uu)$ as $(m,n;\ell,g)$ LRC codes. Two
different codes
of this type may have the same minimum distance $d$, but one of them
may have better $Av_{\rm fail}$ than the other one. This will be shown in the next
example, providing a good illustration of the power of
Theorem~\ref{theo1} over Corollary~\ref{cor1}.

\begin{ex}
\label{ex4}
{\em
Consider two II codes on $4\times n$ arrays with $n>4$ of the same rate
as follows: the first code is a 2-level II
code $\C(n;1,1,1,4)$, and the second code is a 3-level II code
$\C(n;1,1,2,3)$.
According to
Theorem~\ref{theo1}, $\C(n;(1,1,1,4))$ can correct any row with up to 4
erasures as long as the remaining rows do not have more than one
erasure each, and according to Corollary~\ref{cor1}, the minimum distance
of the code is 4. Similarly, $\C(n;(1,1,2,3))$ can
correct one row with up to two erasures, one row with up to three
erasures and up to one erasure in the remaining rows. Its minimum
distance is also 4, by Corollary~\ref{cor1}. However,
by simulation, $Av_{\rm fail}\approx 5.67$  for $\C(n;(1,1,1,4))$ and
$Av_{\rm fail}\approx 6.96$ for $\C(n;(1,1,2,3))$, so the second one is
preferable.

Similarly, consider an optimal LRC code with respect to $4\times n$
arrays, $n>4$, with $\ell\eq 1$ and $g\eq 3$. Hence, it has the same
rate as the two II codes above. According to
bound~(\ref{dist1}), the minimum distance of this
code is 5, better than both II codes. Again by simulation, $Av_{\rm fail}\approx 6.4$, thus, the
optimal LRC code has better $Av_{\rm fail}$ than $\C(n;(1,1,1,4))$
and it is slightly worse than $\C(n;(1,1,2,3))$. If, for instance,
$n\eq 8$, the optimal LRC code
requires a field of size at least 32, while the II codes require a field of
size at least 8.

If we took a PMDS code with the same parameters as the codes above,
we can verify that $Av_{\rm fail}\approx 7.4$. An MDS code with the
same length and dimension has minimum distance 8, so $Av_{\rm fail}\eq 8$, but the
locality is lost.
\qed
}
\end{ex}

Some situations more complex than the ones described in
Example~\ref{ex4} are depicted in Tables~\ref{t1} and~\ref{t2}.
Table~\ref{t1} compares different LRC codes consisting of $16\times 5$
arrays with $\ell\eq 1$ and the number of global parities
$g$ satisfying $3\leq g\leq 8$. For each value of $g$, we compute
$Av_{\rm fail}$ for a PMDS code, for an optimal LRC code, and for an
II $\C(5,\uu)$ code given by a specially selected vector $\uu$. We also
write the value of $Av_{\rm fail}$ for an MDS code with the same
length and dimension, and we have seen that in this case $Av_{\rm fail}$
coincides with the minimum distance of the code. Table~\ref{t2} does
the same thing for $16\times 8$
arrays with $\ell\eq 2$ and the number of global parities
$g\in\{4,5,6,7,12\}$. We can see that in all cases we could find an
II code with larger $Av_{\rm fail}$ than the corresponding optimal
LRC code. Moreover, the II codes in Tables~\ref{t1} and~\ref{t2} may
be implemented over the field
$GF(16)$, while optimal LRC codes require at least the field
$GF(128)$.

\section{Conclusions}
We have presented a method for implementing Integrated Interleaved
codes as Locally Recoverable codes. We
proved the fundamental properties of the codes and we compared their
performance with the one of optimal LRC codes. The main advantage of
II codes is that the fields required in the construction are much
smaller than those of optimal LRC codes. Certainly the minimum
distance of an II code is smaller than the minimum distance of an
optimal LRC code in general (with some exceptions described in this
paper).
However, if we consider the average number
of erasures that an LRC code can tolerate, II codes frequently
outperform optimal LRC codes. PMDS codes maximize
the average number of erasures that an LRC code can tolerate, but their
construction using relatively small fields is an open problem, making
II codes as LRC codes an attractive alternative.




\begin{thebibliography}{99}

\bibitem{ah} K. Abdel-Ghaffar and M. Hassner, ``Multilevel codes for data storage
channels,'' IEEE Trans. on Information Theory, vol. IT-37, pp.~735--41, May
1991.





\bibitem{bgm} M. Blaum, R. M. F. Goodman and R. J. McEliece,
``The Reliability of Single-Error Protected Computer Memories,''
IEEE Trans. on Computers, vol. C-37, pp.~114--19, January 1988.


\bibitem{b}
M. Blaum, ``On Locally Recoverable (LRC) Codes,''
arXiv:1512.06161, December 2015.

\bibitem{bbbmv}
M. Blaum, J. Brady, J. Bruck, J. Menon and A. Vardy,
``The EVENODD Code and its Generalization'' , High Performance Mass
Storage and Parallel I/O: Technologies and Applications, edited by
H. Jin, T. Cortes and R. Buyya,  IEEE \& Wiley  Press, New
York, Chapter 14, pp. 187--208, 2001.

\bibitem{bhh}
M. Blaum, J. L. Hafner and S. R. Hetzler, ``Partial-MDS Codes and their
Application to RAID Type of Architectures,''
IEEE Trans. on Information Theory, vol. IT-59, pp.~4510–-19,
July 2013.


\bibitem{bh}
M. Blaum and S. R. Hetzler, ``Generalized Concatenated Types of Codes
for Erasure Correction,'' arXiv:1406.6270v2, July 2014.


\bibitem{bpsy}
M. Blaum, J. S. Plank, M. Schwartz and E. Yaakobi,
``Partial MDS (PMDS) and Sector-Disk (SD) codes that tolerate the
erasure of two random sectors,''
ISIT 2014,  IEEE International Symposium on Information Theory,
pp.~1792--96, July 2014.


\bibitem{br}
M. Blaum and R. M. Roth, ``New Array Codes for Multiple Phased Burst Correction,"
IEEE Trans. on Information Theory, vol. IT-39,
pp.~66-77, January 1993.

\bibitem{bz}
E. L. Blokh and V. V. Zyablov, ``Coding of Generalized Concatenated
Codes,'' Problemy Peredachii Informatsii, Vol. 10(3), pp.~218--222, 1974.



\bibitem{cm}
V. Cadambe and A. Mazumdar,
``An Upper Bound On the Size of Locally Recoverable Codes,''
International Symposium on Network Coding (NetCod), pp.~1--5, June
2013, also in arXiv:308.13200v2, March 2015.

\bibitem{cw}
J. Campello and B. Wilson, 
``A Generalized Concatenated Scheme: 4K ECC with 8b Symbols,''
unpublished presentation, 2003.





\bibitem{du}
I. Dumer, ``Concatenated Codes and Their Multilevel Generalizations,''
Handbook of Coding Theory, edited by V. S. Pless and W. C.
Huffman, Elsevier Science B. V., Chapter 23, 1998.


\bibitem{g}
G. A. Gibson, ``Redundant Disk Arrays,'' MIT Press, 1992.

\bibitem{gm} R. M. F. Goodman and R. J. McEliece,
``Lifetime analyses of error-control
coded semiconductor RAM systems,'' Proc. IEE, part E, vol. 3, pp.
81--85, 1982.

\bibitem{gm2} R. M. F. Goodman and R. J. McEliece,
``Hamming codes, computer memories, and the birthday
surprise,'' Proc. 20th Allerton Conf. Commun., Control,
Comput., pp.~672--79, 1982.

\bibitem{ghjy} P. Gopalan, C. Huang, B. Jenkins and S. Yekhanin,
``Explicit Maximally Recoverable Codes with Locality,''
IEEE Trans. on Information Theory, vol. IT-60,
pp.~5245--56, September 2014.

\bibitem{ghsy} P. Gopalan, C. Huang, H. Simitci and S. Yekhanin,
``On the Locality of Codeword Symbols,''
IEEE Trans. on Information Theory, vol. IT-58, pp. 6925--34, November
2012.





\bibitem{hl} J. Han and L. A. Lastras-Monta\~no, ``Reliable Memories
with Subline Accesses,'' ISIT 2007, IEEE International Symposium on
Information Theory, pp.~2531--35, June 2007.

\bibitem{hapkt}
M. Hassner, K. Abdel-Ghaffar, A. Patel, R. Koetter and B. Trager,
``Integrated Interleaving -- A Novel ECC Architecture,'' IEEE
Transactions on Magnetics, Vol. 37, No. 2, pp.~773--5, March 2001.

\bibitem{hcl}
C. Huang, M. Chen and J. Li, ``Pyramid Codes: Flexible Schemes to
Trade Space for Access Efficiency in Reliable Data Storage Systems,''
ACM Transactions on Storage, Vol. 9, No. 1, Article 3, March 2013.

\bibitem{hsx}
C. Huang, H. Simitci, Y. Xu, A. Ogus, B. Calder, P. Gopalan, J. Li
and S. Yekhanin,
``Erasure Coding in Windows Azure Storage,'' 2012 USENIX Annual
Technical Conference, Boston, Massachussetts, June 2012.

\bibitem{kn}
M. S. Klamkin and D. J. Newman, ``Extensions of the Birthday
Surprise,'' J. Combin. Theory 3, pp.~279--82, 1967.

\bibitem{kna}
M. Kuijper and D. Napp, ``Erasure codes with simplex locality,''
arXiv:1403.2779, March 2014.

\bibitem{ll}
M. Li and P. C. Lee,
``STAIR Codes: A General Family of Erasure Codes
for Tolerating Device and Sector Failures in
Practical Storage Systems,''
12th USENIX Conference on File and Storage Technologies (FAST '14),
Santa Clara, CA, February 2014.













\bibitem{ms} F. J. MacWilliams and N. J. A. Sloane, ``The Theory
of Error-Correcting Codes,'' North Holland, Amsterdam, 1977.


\bibitem{me}
L. Mearian, ``SSDs do die, as Linus Torvalds just discovered,''
http://www.computerworld.com/article/2484998/solid-state-drives/ssds-do-die---as-linus-torvalds-just-discovered.html.

\bibitem{M}
Micron, ``TN-29-17: NAND Flash Design and Use Considerations
Introduction,'' Micron Technology, Inc., 2006.


\bibitem{pd}
D. S. Papailiopoulos and A. G. Dimakis,
``Locally Repairable Codes,'' IEEE Trans. on Information Theory, vol.
IT-60, pp.~5843–-55,
October 2014.

\bibitem{pa}
A. Patel, ``Two-Level Coding for Error-Control in Magnetic Disk
Storage Products,'' IBM Journal of Research and Development, vol. 33,
pp.~470--84, 1989.



\bibitem{pb} J. S. Plank and M. Blaum, ``Sector-Disk (SD) Erasure Codes for Mixed Failure Modes
in RAID Systems,'' ACM Transactions on Storage,
Vol. 10, No. 1, Article 4, January 2014.

\bibitem{pbh} J. S. Plank, M. Blaum and J. L. Hafner, ``SD Codes:
Erasure Codes Designed for How Storage Systems
Really Fail,'' 11th USENIX
Conference on File and Storage Technologies (FAST '13),
Santa Clara, CA, February 2013.

\bibitem{pklk}
N. Prakash, G. M. Kamath, V. Lalitha and P. V. Kumar, ``Optimal
linear codes with a local-error-correction property,'' 
ISIT 2012, IEEE International Symposium on
Information Theory, pp.~2776--80, July 2012, and 
arXiv:1202.2414, February 2012.


\bibitem{rk}
A. S. Rawat, O. O. Koyluoglu, N. Silberstein and S. Vishwanath,
``Optimal Locally Repairable and Secure Codes for Distributed Storage Systems,''
IEEE Trans. on Information Theory, vol. IT-60, pp.~212–-36,
January 2014.


\bibitem{sa}
M. Sathiamoorthy, M. Asteris, D. Papailiopoulos, A. G. Dimakis, R.
Vadali, S. Chen and D. Borthakur,
``XORing Elephants: Novel Erasure Codes for Big Data,''
 Proceedings of VLDB, Vol.~6, No.~5, pp.~325--336, August 2013.

\bibitem{sd}
W. Song, S. H. Dau, C. Yuen and T. J. Li,
``Optimal Locally Repairable Linear Codes,''
IEEE Journal on Selected Areas in Communications,
Vol.~32 , pp.~1019--36, May 2014.

\bibitem{tb} I. Tamo and A. Barg, ``A Family of Optimal Locally
Recoverable Codes,'' IEEE Trans. on Information Theory, vol. IT-60,
pp.~4661--76, August 2014.

\bibitem{twb} I. Tamo, Z. Wang and J. Bruck,
``Zigzag Codes: MDS Array Codes With Optimal Rebuilding,''
IEEE Trans. on Information Theory, vol. IT-59, pp.~1597--616, March 2013.

\bibitem{tk} X. Tang and R. Koetter, ``A Novel Method for Combining
Algebraic Decoding and Iterative Processing,''
ISIT 2006, IEEE International Symposium on Information Theory,
pp.~474--78, July 2006.










\bibitem{wz}
A. Wang and Z. Zhang,
``Repair Locality with Multiple Erasure Tolerance,''
IEEE Trans. on Information Theory, vol. IT-60, pp.~6979--87, November 2014.

\bibitem{w}
Y. Wu, ``Generalized integrated interleaving codes,'' submitted for
publication, 2015.

\bibitem{zy}
A. Zeh and E. Yaakobi, ``Bounds and Constructions of Codes with
Multiple Localities,''  arXiv:1601.02763, January 2016.

\bibitem{z}
V. A. Zinoviev, ``Generalized cascade codes,'' Probl. Pered. Inform.,
vol. 12, no. 1, pp.~5–-15, 1976.

\end{thebibliography}
\end{document}